\documentclass{interact}

\usepackage{epstopdf}% To incorporate .eps illustrations using PDFLaTeX, etc.
\usepackage[caption=false]{subfig}% Support for small, `sub' figures and tables
\usepackage{xcolor}

\usepackage[numbers,sort&compress]{natbib}% Citation support using natbib.sty
\bibpunct[, ]{[}{]}{,}{n}{,}{,}% Citation support using natbib.sty
\theoremstyle{plain}% Theorem-like structures provided by amsthm.sty

\theoremstyle{definition}

\theoremstyle{remark}

\begin{document}

%\articletype{ARTICLE TEMPLATE}% Specify the article type or omit as appropriate

\title{Statistical modeling of breast cancer radiomic features and hazard using image registration-aided longitudinal CT data}

\author{
\name{Subrata Mukherjee\textsuperscript{a,*}, Qian Cao\textsuperscript{a} \thanks{CONTACT Qian Cao. Email: Qian.Cao@fda.hhs.gov}, Thibaud Coroller\textsuperscript{b}, Ravi K. Samala\textsuperscript{a}, Nicholas Petrick\textsuperscript{a} and Berkman Sahiner\textsuperscript{a}}
\affil{\textsuperscript{a} Center for Devices and Radiological Health, U.S. Food and Drug Administration, \\Silver Spring, MD \\
\textsuperscript{b} Novartis Pharmaceuticals Corp, Hanover, NJ}
}

% \title{Statistical modeling of breast cancer radiomic features and hazard using image registration-aided longitudinal CT data}

% \author{
% \name{Subrata Mukherjee\textsuperscript{a}, Qian Cao\textsuperscript{b} \thanks{CONTACT Qian Cao. Email: Qian.Cao@fda.hhs.gov}, Thibaud Coroller\textsuperscript{c}, Ravi K. Samala\textsuperscript{b},  Nicholas Petrick\textsuperscript{b} and Berkman Sahiner\textsuperscript{b} }
% \affil{\textsuperscript{a} Energy Science \& Technology Directorate, Sensors \& Electronics Group, Oak Ridge National Laboratory, \\Oak Ridge, TN \\
% \textsuperscript{b} Center for Devices and Radiological Health, U.S. Food and Drug Administration, \\Silver Spring, MD \\
% \textsuperscript{c}, Novartis Pharmaceuticals Corp \\Hanover, NJ}
% }

\maketitle

\begingroup
\renewcommand\thefootnote{*}
\footnotetext{Now at the Energy Science \& Technology Directorate, Sensors \& Electronics Group, Oak Ridge National Laboratory, Oak Ridge, TN}
\endgroup

\begin{abstract}
Patients with metastatic breast cancer (mBC) undergo repeated computed tomography (CT) imaging during treatment to monitor disease progression. Accurate longitudinal tracking of individual lesions across scans from multiple radiologists is essential for reliable radiomic analysis and clinical decision-making. We conducted a retrospective study using serial chest CT scans from the Phase III MONALEESA-3 and MONALEESA-7 trials and developed statistical models for multi-source data integration and survival analysis. First, we introduced a Registration-based Automated Matching and Correspondence (RAMAC) algorithm to establish lesion correspondence across annotations from different radiologists and imaging time points using the Hungarian algorithm. Second, using the RAMAC-processed dataset, we developed interpretable radiomic survival models for progression-free survival prediction by combining baseline radiomic features, post-treatment changes at Weeks 8, 16, and 24, and demographic variables. To address the high dimensionality of longitudinal radiomic data, feature reduction was performed using an L1-penalized additive Cox proportional hazards model and best subset selection followed by Cox modeling. Model performance was evaluated using the concordance index (C-index). Incorporating additional imaging time points improved predictive performance, increasing the mean C-index from 0.58 at baseline to 0.64. Joint modeling further showed significant associations between longitudinal radiomic features and survival outcomes over time.
\end{abstract}

\begin{keywords}
Survival modeling; joint modeling; feature selection; registration; lesion correspondence; non-invasive imaging
\end{keywords}

\section{Introduction}

This paper presents an ongoing collaboration between the U.S. Food and Drug Administration (FDA) and Novartis, utilizing anonymized data from two Phase III clinical trials, MONALEESA-3 and MONALEESA-7, to assess the impact of the investigational drug ribociclib \cite{coroller2024methodology}. These randomized, double-blind trials aimed to evaluate the efficacy of ribociclib in combination with endocrine therapy (ET) for the treatment of metastatic breast cancer (mBC), specifically in patients with hormone receptor-positive (HR+) and human epidermal growth factor receptor 2-negative (HER2-) status \cite{tripathy2018ribociclib,slamon2018phase,hortobagyi2018ribociclib}. Metastasis remains a leading cause of cancer-related mortality \cite{chaffer2011perspective}, necessitating the use of longitudinal imaging for accurate diagnosis and treatment monitoring.

In our retrospective study, we employed computed tomography (CT) scan images, a non-invasive imaging modality, acquired at multiple time points to evaluate tumor progression and inform treatment decisions. Tumor evaluations were conducted at baseline (screening), followed by every 8 weeks during the first 18 months, and then every 12 weeks until disease progression, patient death, or withdrawal from the study. We focused on the first four time points: Screening/Baseline, Week 8, Week 16, and Week 24. Metastatic target lesions (mBC lesions that are actively monitored and tracked) were annotated by either a radiologist from one radiologist group, or two independent radiologists from two groups resulting in multi-source radiological annotations. The radiologists adhered to Response Evaluation Criteria in Solid Tumors (RECIST) v1.1. Two-dimensional (2D) segmentations of RECIST target lesions were also provided by the same radiologists. However, the computerized analysis of this longitudinal CT data presents challenges, including the appearance of new lesions, disappearance of responding lesions, inter radiologist variability in multi-source annotations across time points, and variations in the field of view (FOV) due to different scan series \cite{humbert2020dissociated,wang2012study,huff2023performance}.

To address these challenges, we developed an automated lesion correspondence algorithm, Registration based Automated Matching and Correspondence (RAMAC) \cite{10627649}, capable of effectively tracking distinct lesions throughout the longitudinal imaging study. After establishing lesion correspondence, we computed radiomic features for each target lesion at baseline and the subsequent three time points. We then developed radiomic-based survival models to predict the risk of cancer progression in patients, using progression-free survival (PFS) as the endpoint. The survival data included the time of PFS failure or censoring and a status indicator for event occurrence. We list our contributions below:

\begin{enumerate}
    \item 
    To understand how changes in various radiomic features over time impact disease progression and treatment effects, it is important to consistently track the lesions from which the radiomic features are derived. However, this task isvchallenging due to the multi-source nature of the data, where lesions are annotated by different radiologists at different time points, and lesion labeling may not be consistent across radiologists. Efficiently integrating data from multiple radiologists and across different time frames is needed for ensuring consistency and reliability in subsequent analyses.
In RAMAC, we develop an unsupervised, automated, and robust framework for integrating multi-source radiomic data \cite{10627649,Mukherjee2024RAMAC}. We then demonstrate how downstream statistical analyses based on RAMAC-integrated data can be effectively conducted, as discussed in the following two points.
Overall, this approach provides a statistical framework for analyzing radiomic feature importance, treatment effects, and temporal changes in radiomic features by integrating multi-source CT scan data, which is often encountered in real-world clinical applications \cite{de2020multi}.

    \item For each target lesion, 98 radiomic features were computed at baseline, at the first post-treatment inspection at Week 8, then at Week 16 and Week 24, respectively. To address the challenge of high-dimensional data, where the number of radiomic features across columns (timepoint-specific covariates) exceeds the number of patient records (rows), we employed two feature reduction techniques:
    \begin{enumerate}
        \item First, we applied a regularized additive Cox proportional hazards model \cite{tay2023regularized}, utilizing 5-fold cross-validation (CV). This process was conducted separately for Baseline, Baseline combined with delta radiomic features from Week 8, Baseline with delta radiomic from Week 16 and Week 8, and finally, Baseline with delta radiomic from Week 8, Week 16, and Week 24. Model performance was assessed using the concordance index (C-index) on the test data.
        \item We implemented variable selection using Best Subset Selection (BESS) \cite{wen2020bess,peterson2018variable} to identify the top four radiomic features from each time point separately. An additive Cox proportional hazards model \cite{therneau1997extending,rod2012additive} was subsequently fitted using the reduced feature set, and model performance was again evaluated using the C-index on the test dataset.
    \end{enumerate}

    \item A comprehensive joint modeling framework was introduced to explore the time-dependent associations between longitudinal radiomic features and survival outcomes. This approach allowed us to investigate the association between radiomic features and survival, quantifying the benefit of using longitudinal radiomic data compared to baseline radiomic alone.
\end{enumerate}

\begin{figure}[!htbp]
\centering
  \includegraphics[width=1\textwidth]{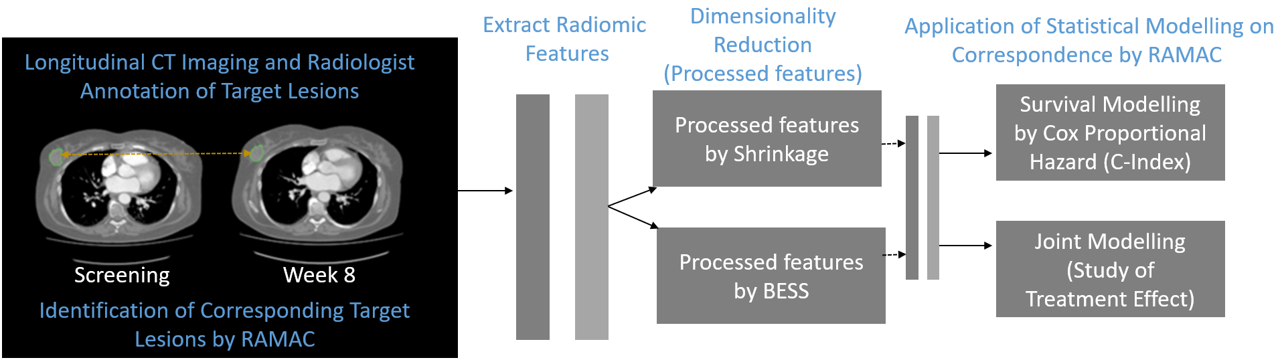}
  \caption{Overview of this study}\label{fig1}
\end{figure}

Figure~\ref{fig1} provides a schematic overview of the entire framework employed in this study. The remainder of the paper is organized as follows: Section~\ref{datagen} details the organization of structured data by RAMAC across various scanning timepoints; Section~\ref{featureprocess} discusses the various radiomic and clinical features extracted from the structured data. In section~\ref{shrinkage} and section~\ref{bess} we studied reduced feature set by shrinkage and best subset selection based reduction techniques. On the reduced dataset various survival analysis models based on the additive Cox proportional hazards model and their outcomes are studied; Section~\ref{jm} shows the joint modeling of radiomic features; and Section~\ref{conclusion} concludes the study with a discussion and potential future work.

\section{Data Preprocessing}\label{datagen}
One major challenge with the dataset was the inconsistent labeling of lesion (e.g. lesion $G_1$ at baseline but labeled lesion $G_2$ at follow-up). At each timepoint, there may be one or more image series annotated by one or two radiologist groups, with annotations of target lesions varying across the dataset. To sort out the consistency and match lesion over time, the RAMAC algorithm was developed in \cite{10627649} to establish correspondence in lesion centroids. 

\begin{figure}[!htbp]
\centering
  \includegraphics[width=1\textwidth]{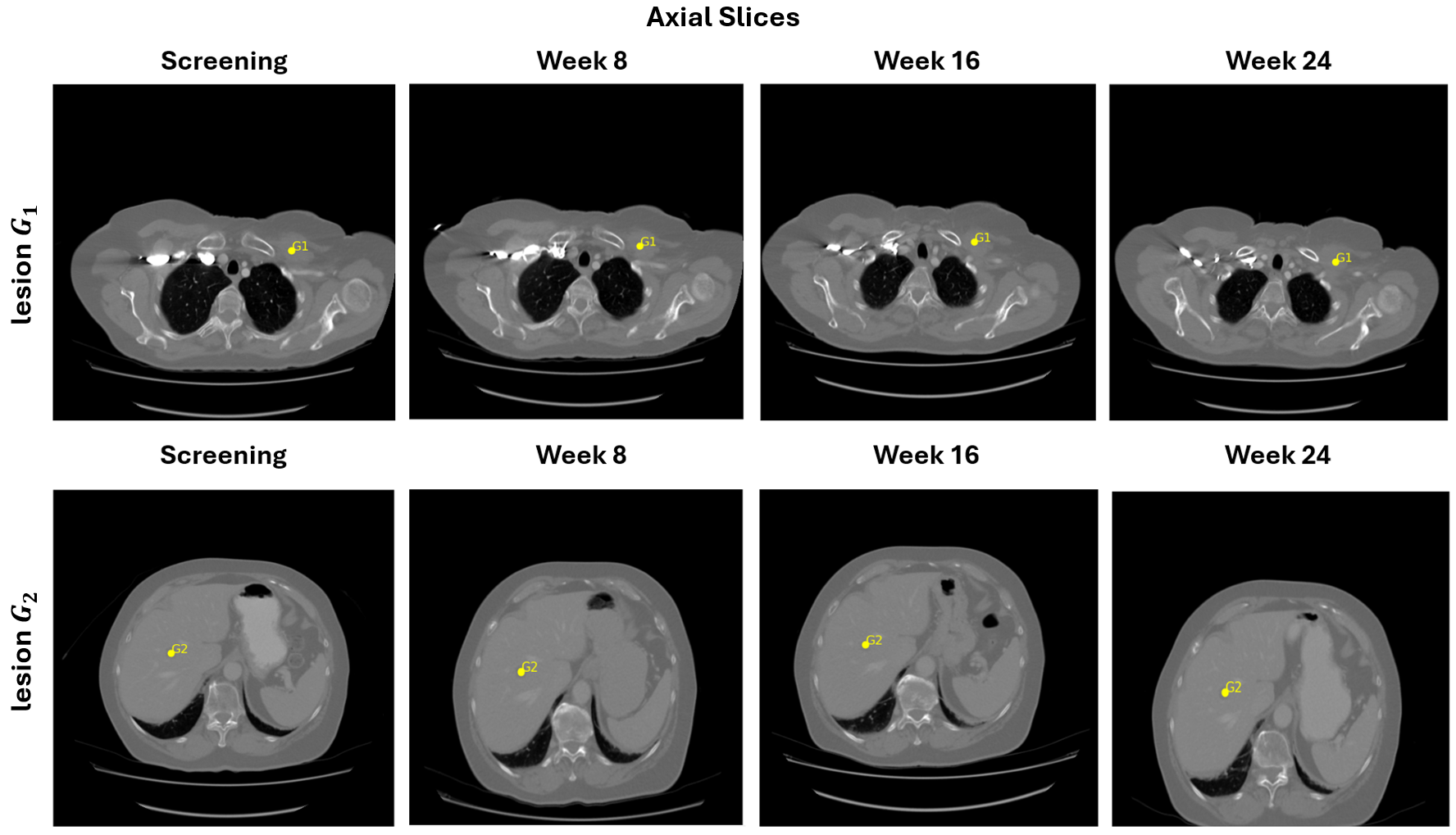}
  \caption{Visualization of RAMAC algorithm in identifying and tracking lesions across different timepoints for one patient in the data pipeline (lesion centroids marked in yellow, upper row represents the G\textsubscript{1} lesion across timepoints and bottom row represents the G\textsubscript{2} lesion across timepoints)}\label{fig2}
\end{figure}

RAMAC operates in a two-phase pipeline. In the first phase, the Simple Image Toolkit (SITK) framework \cite{marstal2016simpleelastix} is used to perform volumetric rigid registration \cite{an2021robust,marstal2016simpleelastix}, aligning different image series from the same time point or across different time points. The registration framework requires two DICOM image series and lesion coordinates (landmarks or ROIs) in physical coordinates as inputs where the image series designated as the 'fixed' image serves as the reference, while the 'moving' image is transformed to align with the fixed image. In the second phase, the Hungarian algorithm \cite{munkres1957algorithms,mills2007dynamic,larrey2006optimal}, based on a distance measure, is used to establish correspondence between the transformed moving coordinates (lesion centroids in the moving image) and the fixed lesion coordinates (lesion coordinates of the fixed image). The Hungarian algorithm finds one-to-one correspondence between two lesion sets by minimizing the Euclidean distance between lesion centroids in the two sets. The alignment of registered and fixed images post-registration is visually demonstrated through triaxial plots (Axial Slice, Coronal Slice, Sagittal Slice, Lesion ROI), with lesion centroids highlighted by yellow dots. Figure~\ref{fig2} (axial slices) shows the consistency in the spatial positioning of two target lesion (G\textsubscript{1}, G\textsubscript{2}) across various timepoints (Screening, Week 8, Week 16, and Week 24) for one such patient. This implementation is publicly available \cite{Mukherjee2024RAMAC}.

\section{Radiomic Feature Extraction and Analysis}\label{featureprocess}

For each CT 2D slice containing a RECIST target lesion (i.e., each image), 98 radiomic features were computed at four inspection time points: baseline, Week 8, Week 16, and Week 24 of post treatment inspection. These 98 radiomic features included 9 two-dimensional shape-based features, 15 first-order statistical features, 23 gray-level co-occurrence matrix (GLCM) features, 16 gray-level run length matrix (GLRLM) features, 16 gray-level size zone matrix (GLSZM) features, 5 neighboring gray-tone difference matrix (NGTDM) features, and 14 gray-level dependence matrix (GLDM) features. The extraction of these radiomic features from the registered volumetric CT images was performed using the Pyradiomic package [28].

\begin{figure}[!htbp]
\centering
  \includegraphics[width=0.8\textwidth]{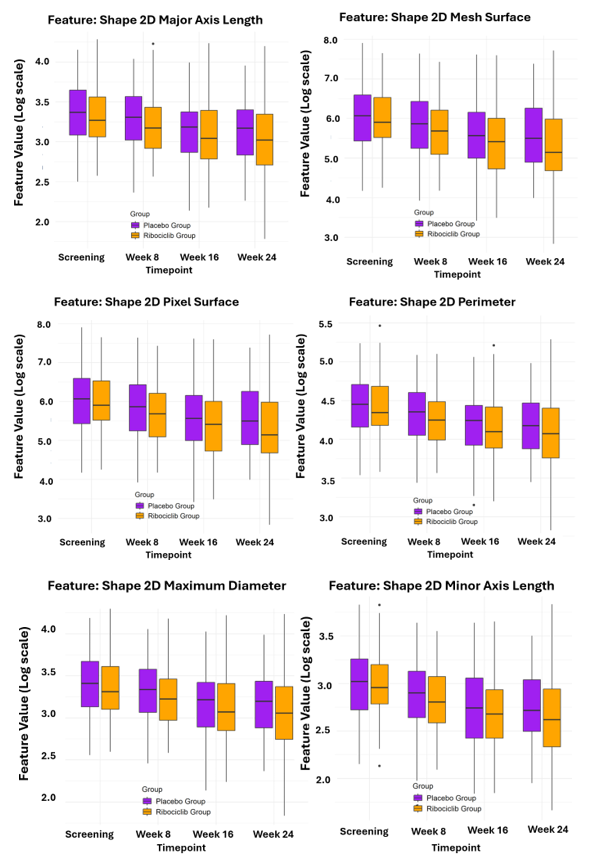}
  \caption{Box plot the 2D radiomic shape features in log scale based on training cohort across time points stratified by group (ribiociclib in orange and placebo in purple). The median is demarcated by the black line and the lower and upper edges of the boxes correspond to the 25th (Q1) and 75th (Q3) percentiles }\label{fig3}
\end{figure}

To assess the behavior of the 2D radiomic shape features across the four time points, we employed box plots as shown in figure~\ref{fig3} where the feature values in log scale were plotted against timepoints. These model free analyses and visualizations compare the distribution of each feature within the training dataset across the four time points and provide visualizations of each feature  separately, stratified into two groups: ribociclib (depicted in orange) and placebo (depicted in purple). The central line in the plots represents the median value of each feature, while the lower and upper edges of the box plot correspond to the 25th (Q1) and 75th (Q3) percentiles, thereby defining the interquartile range. The visualizations reveal a distinct trend, specifically the median level in the treatment group starting from Week 8 to be usually below the median level in the placebo group.

Delta radiomic features at each post-screening time point were also calculated as the ratio of the radiomic feature values at that specific time point to those same values at  screening baseline. For example, the delta radiomic values for Week 8 are the ratio of each feature value at Week 8 to its corresponding value at Screening. Similarly, for Week 16, and week 24.

\section{Survival Modeling with Shrinkage Models}\label{shrinkage}
We utilized CT data from 203 breast cancer patients enrolled in the clinical trial, with target lesions annotated by radiologists. The dataset was divided into training and testing cohorts, comprising 111 and 92 patients, respectively.

For each target lesion, 98 radiomic features were extracted at baseline, and delta radiomic features were calculated at Week 8, Week 16, and Week 24, yielding a total of 392 radiomic features per patient. These radiomic features were combined with 9 demographic variables, resulting in a total of 401 features used to predict progression-free survival (PFS) for patients in the clinical trial. Notably, 55\% of the patients had censored outcomes due to loss to follow-up or absence of an event. The 9 demographic variables included LIVERBL (Liver Involvement), PRENGR1 (Prior Endocrine Therapy), ENDSNBL (Endocrine Therapy Status), AST (Aspartate Aminotransferase), ALP (Alkaline Phosphatase), Age, Denovo (De Novo Disease), HGB (Hemoglobin), and ARM (Treatment Arm). For each timepoint, if multiple target lesions were present for a patient, the feature vectors were averaged across all target lesions.

Subsequently, an additive Cox proportional hazards model was fitted on the training data to predict lesion progression risk based on the aggregated patient-level feature vectors. It was assumed that treatment allocation was random, ensuring no confounding with baseline screening observations $(\mathbf{x}_i, \mathbf{U}_i)$, where Treatment Arm $(T_i)$ is independent of $(\mathbf{x}_i, \mathbf{U}_i)$ and any other unobserved variables that could potentially influence the outcome. Here, $\mathbf{x}_i$ represents the vector of radiomic features at Screening, $\mathbf{U}_i$ represents time-invariant demographic features, and $T_i$ denotes the treatment arm (ribociclib vs. placebo) for a particular patient $i$.

Let $\mathbf{x}_i$ be the baseline feature vector for a subject $i$, $\mathbf{Z}_i$ is the vector of changes observed at Week 8 compared to Screening, $\mathbf{W}_i$ is the vector of changes observed at Week 16 compared to Screening, and $\mathbf{V}_i$ is the vector of changes observed at Week 24 compared to Screening. The hazard function at time $t$ is given by:

\[
\log \left( \frac{h(t \mid \mathbf{x}_i, \mathbf{Z}_i, \mathbf{W}_i, \mathbf{V}_i, \mathbf{U}_i, T_i)}{h_0(t)} \right)
= f_1(\mathbf{x}_i) + f_2(\mathbf{Z}_i) + f_3(\mathbf{W}_i) + f_4(\mathbf{V}_i) + f_5(\mathbf{U}_i) + T_i,
\]
where $h_0(t)$ is the baseline hazard function that is independent of covariates, and $f_1, f_2, f_3, f_4, f_5$ are additive functions defined as:
\[
f_1(\mathbf{x}_i) = \mathbf{x}_i' \boldsymbol{\beta}, \quad
f_2(\mathbf{Z}_i) = \mathbf{Z}_i' \boldsymbol{\gamma}, \quad
f_3(\mathbf{W}_i) = \mathbf{W}_i' \boldsymbol{\rho}, \quad
f_4(\mathbf{V}_i) = \mathbf{V}_i' \boldsymbol{\alpha}, \quad
f_5(\mathbf{U}_i) = \mathbf{U}_i' \boldsymbol{\mu}.
\]

$\boldsymbol{\beta}, \boldsymbol{\gamma}, \boldsymbol{\rho}, \boldsymbol{\alpha}, \boldsymbol{\mu}$ are the coefficient vectors associated with each set of covariates. The transposed feature vectors, ensuring proper matrix multiplication with the corresponding coefficients, are denoted as $\mathbf{x}_i', \mathbf{Z}_i', \mathbf{W}_i', \mathbf{V}_i', \mathbf{U}_i'$.

\begin{figure}[!htbp]
\centering
  \includegraphics[width=1\textwidth]{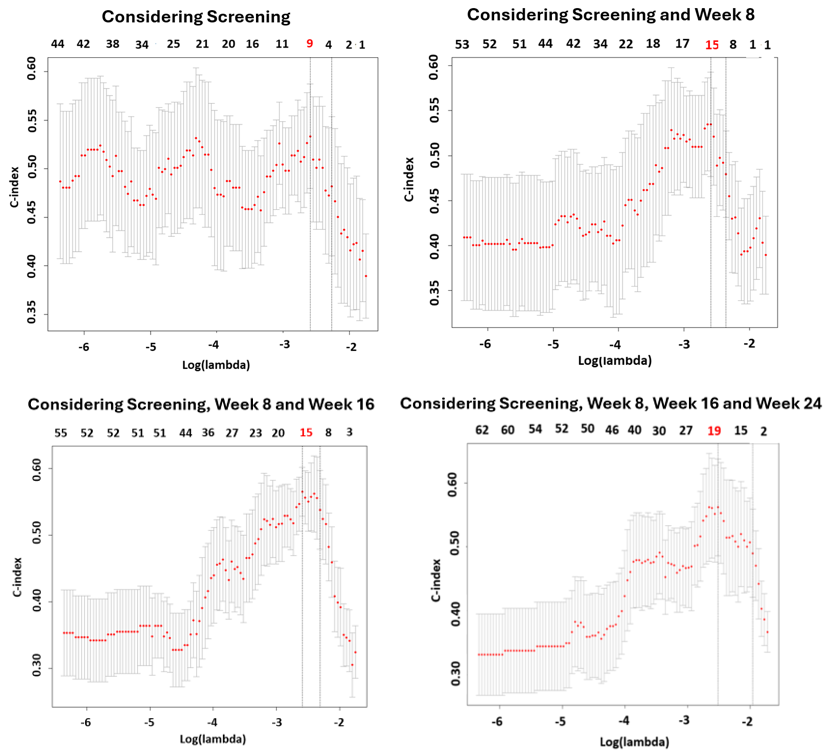}
  \caption{Cross-validation (CV) paths illustrating the relationship between the regularization parameter $(\lambda)$, the Harrell Concordance Index (C-index), and the number of non-zero coefficients based on training data. Each plot represents a different feature set: the upper left plot considers only Baseline features (Screening), the upper right plot incorporates two time points (Screening and Week 8), the lower left plot includes three time points (Screening, Week 8, and Week 16) and the lower right includes four time points (Screening, Week 8, Week 16, and Week 24). The x-axis denotes log-transformed $\lambda$ values, while the y-axis represents the C-index. The number of non-zero coefficients is annotated at the top of each plot (one marked in red shows the non zero coefficients against the optimum $\lambda$), with red dots indicating the mean C-index across cross-validation folds and error bars representing variability.}\label{fig4}
\end{figure}

Given the high dimensionality of the data, the vectors $\mathbf{x}_i, \mathbf{Z}_i, \mathbf{W}_i, \mathbf{V}_i, \mathbf{U}_i$ were reduced using a regularized Cox regression model applied to the training data utilizing the \texttt{glmnet} package \cite{tay2023regularized}. In \texttt{glmnet}, the negative log of the partial likelihood is penalized using an elastic net penalty ($L_{1}$). Let $y_i$ be the observed event time (or censoring time) for patient i. The likelihood that patient $i$ died at time $Y_i$ is given by:
\[
L_i(\boldsymbol{\theta}) =
\frac{h(y_i \mid \text{data}_i)}{
\sum\limits_{j: y_j \geq y_i} h(y_i \mid \text{data}_j)},
\]
where $\text{data}_j = [\mathbf{x}_j; \mathbf{Z}_j; \mathbf{W}_j; \mathbf{V}_j; \mathbf{U}_j; T_j]$ and
$\boldsymbol{\theta} = [\boldsymbol{\beta}; \boldsymbol{\gamma}; \boldsymbol{\rho}; \boldsymbol{\alpha}; \boldsymbol{\mu}]$.
This can be rewritten as:
\[
L_i(\boldsymbol{\theta}) =
\frac{h_0(y_i) \exp\{f(\text{data}_i)\}}{
\sum\limits_{j: y_j \geq y_i} h_0(y_i) \exp\{f(\text{data}_j)\}},
\]
where $f = f_1 + f_2 + f_3 + f_4 + f_5 + T$.
Thus, it reduces to $
L_i(\boldsymbol{\theta}) = {\exp\{f(\text{data}_i)\}}({\sum \exp\{f(\text{data}_j)\}})^{-1}$, where, 
the sum in the denominator is over all such $j$ where patient $j$ did not die before patient $i$. A L1-penalized Cox proportional hazards model was fitted using 5-fold cross-validation (CV) separately for (1) Baseline, (2) Baseline with delta radiomic from Week 8, (3) Baseline with delta radiomic from Week 8 and Week 16, and (4) Baseline with delta radiomic from Week 8, Week 16, and Week 24. The cross-validation paths (CV paths) in Figure~\ref{fig4} illustrate the C-index and the number of non-zero coefficients as a function of the regularization parameter $\lambda$.

5-fold cross-validation was implemented using the cv.glmnet function \cite{tay2023regularized}, which systematically partitions the dataset into training and validation subsets. At each iteration, the model is trained on four folds, while the remaining fold serves as the validation set. This process is repeated five times, ensuring that each fold is used for validation once. The Harrell Concordance Index (C-index) is computed for each fold and subsequently averaged across all five folds to provide an aggregated estimate of predictive performance for each value of $\lambda$.

Regularization was applied using the $L_{1}$-norm penalty (lasso), which promotes sparsity by shrinking some coefficients to zero, thereby facilitating feature selection. The penalization strength is controlled by $\lambda$, where higher values induce stronger regularization, reducing the number of non-zero coefficients and simplifying the model. The optimal $\lambda$ is determined as the value that maximizes the mean C-index across all CV folds.

The CV paths are shown in Figure~\ref{fig4} where a CV path is a graphical representation of model performance and feature selection stability across different values of $\lambda$. In Figure~\ref{fig4}, each plot presents the C-index trajectory (y-axis) as a function of log-transformed $\lambda$ values (x-axis), with the number of non-zero coefficients displayed at the top. The red dots indicate mean C-index values across cross-validation folds, while error bars represent variability. The observed trend demonstrates that the inclusion of longitudinal delta radiomic features enhances model performance.

\subsection{Shrinkage Model Results}\label{shrinkageresults}

The C-index was used to evaluate the performance of the fitted regularized Cox model on the test data based on the features obtained from the previous steps. The C-index provides a robust metric for assessing model performance by measuring its ability to correctly rank the relative risk or survival times of patients at different time points. When considering four time points, the model included 19 radiomic features and 9 demographic features, resulting in a total of 28 features. For three and two time points, the model included 15 radiomic and 9 demographic features, totaling 24 features each. For the baseline scenario, there were 9 radiomic and 9 demographic features, totaling 18 features.

To quantify model performance, the C-index was computed for each model configuration on the test dataset, and 95\% confidence intervals (CIs) were estimated using a bootstrapping approach. Specifically, 1,000 bootstrap resamples were generated by sampling the test dataset with replacement, and the C-index was recalculated for each resample. The mean C-index and its standard error were computed from the bootstrap distribution, and the 95\% confidence interval was derived using the standard normal approximation: \[
\text{95\% CI} = \bar{C} \pm 1.96 \times SE_C
\]
where \(\bar{C}\) is the mean bootstrapped {C-index}, and \(SE_C\) is its standard error.

\begin{figure}[!h]
    \centering
    \includegraphics[width=0.8\textwidth, height = 4in ]{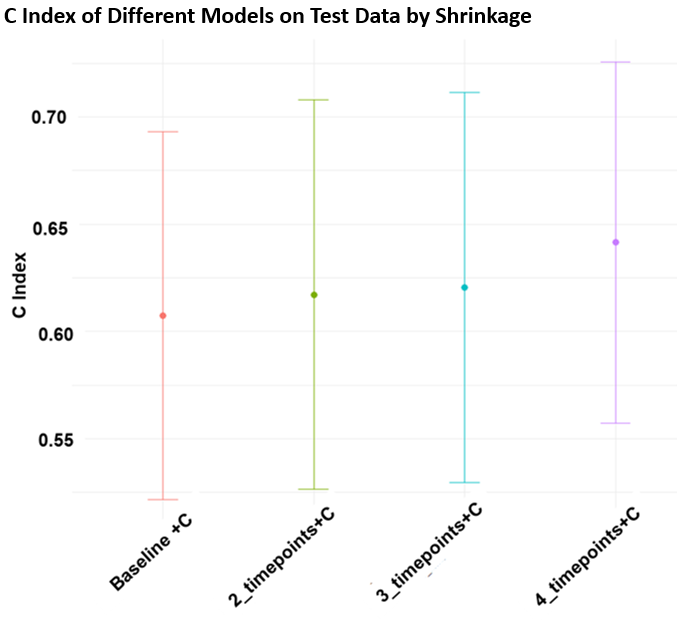} 
    \caption{Performance of shrinkage‐based Cox models on the test dataset across different numbers of longitudinal time points. “C” denotes inclusion of demographic features. Error bars represent the 95\% bootstrap confidence intervals, obtained from 1,000 resamples of the test dataset. The figure illustrates the improvement in the C-index as additional time points are incorporated into the model.}
    \label{fig5}
\end{figure}

Figure~\ref{fig5} presents the test set performance of the Cox models at different time points. The mean C-index increased from 0.6073 for the baseline model (Screening only) to 0.6171 when incorporating two time points (Screening and Week 8). Further improvements were observed with three time points (Screening, Week 8, and Week 16), yielding a C-index of 0.6204. The best model performance was achieved with four time points (Screening, Week 8, Week 16, and Week 24), resulting in a C-index of 0.6415 with a 95\% CI of (0.557, 0.726). The models with different timepoints are statistically significant over the baseline model based on the Z-test \cite{kang2015comparing}.

\section{Survival Modeling with Best Subset Selection-based Variable Selection}\label{bess}
In addition to shrinkage-based variable selection, we implemented variable selection using Best Subset Selection (BESS) \cite{wen2020bess, peterson2018variable}, a statistical approach that identifies the optimal subset of predictors. BESS enhances model performance by selecting a combination of features that best explain the outcome while reducing the risk of overfitting by excluding less informative variables. We applied subset selection via BESS sequentially to features from each timepoint, selecting the top 4 features from each section separately. Specifically, the top 4 features were selected from radiomic features at Screening, followed by the top 4 delta radiomcs features from Week 8, from Week 16, and finally from Week 24.

Considering the Screening timepoint alone, this approach resulted in a total of 13 features: 4 radiomic features from Screening and 9 demographic variables. When extending to the Week 8 timepoint, we added the top 4 Week 8 delta radiomic yielding 17 features in total. Similarly, when extending to the  Week 16 and Week 24 timepoints, which included a total of 21 and 25 features, respectively..

Cox proportional hazards models were fitted separately to the training dataset for each set of time points, and the C-index was used to evaluate the performance of the fitted Cox models on the test data, based on the selected features.

\subsection{Best Subset Selection Model Results}\label{bessresults}
\begin{figure}[!h]
    \centering
    \includegraphics[width=0.8\textwidth,height = 4 in]{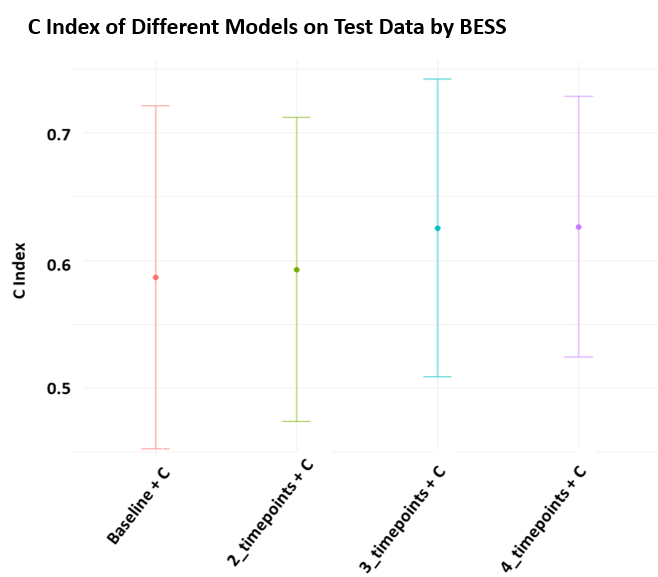} 
    \caption{Performance of Cox proportional hazards models on the test data using BESS for selecting the top 4 radiomic features."C" represents the inclusion of demographic features. Plots illustrate the increase in the C-index of the Cox models as additional time points are incorporated.}
    \label{fig6}
\end{figure}

As shown in Figure~\ref{fig6}, there was a steady increase in the C-index as additional time points were included in the model. The mean C-index improved from 0.5869 when only the Screening timepoint was used, to 0.5930 when time points up to Week 8 were included. Incorporating time points up to Week 16 further increased the mean C-index to 0.6252. When time points up to Week 24 were considered, the mean C-index rose to 0.6262 (95\% CI: 0.524–0.728).

\section{Joint Modeling Analysis}\label{jm}

Unlike the methods used in section~\ref{shrinkage} and section~\ref{bess}, which either consider separate additive effects of the radiomic features across time points or sequentially consider their effects separately, we now consider the effect of these radiomic features jointly. Joint modeling (JM) is an advanced statistical methodology \cite{therneau1997extending,rod2012additive,zhudenkov2022workflow,rizopoulos2012joint,rizopoulos2024using,andrinopoulou2021reflection} that facilitates the investigation of clinical trial outcomes by quantifying the association between baseline and/or longitudinal radiomic features and event risk. This approach belongs to a class of parametric survival models, often based on proportional hazards models, which enable the integration of both baseline (screening) and dynamic longitudinal radiomic data (radiomic obtained at subsequent time points). 

\noindent Traditional time-dependent additive Cox models, as discussed earlier, face several limitations: (a) Time-varying radiomic covariates are endogenous \cite{therneau1997extending,rod2012additive}, meaning they are influenced by other variables within the system being modeled. Importantly, changes in radiomic features at subsequent time points are dependent on the treatment arm. (b) Including additional time points in the analysis often leads to a reduction in the number of patients available for inclusion due to censoring or missing data. Joint modeling overcomes these limitations by providing a comprehensive framework for exploring longitudinal data, offering insights into the relationships between dynamic radiomic features and survival outcomes, and potentially leading to robust survival models.

\noindent The multivariate joint model analysis consists of three components: 
(1) Longitudinal Component, (2) Survival Component, and (3) Joint Component.

Each longitudinal radiomic feature is modeled using a linear mixed-effects model (LME) \cite{zhudenkov2022workflow}. Let $y_{ij}^{(l)}$ denote the magnitude of the $l^{\text{th}}$ radiomic feature for the $i^{\text{th}}$ patient recorded at the $j^{th}$ observation point and $t_{ij}$ be the corresponding time point. The longitudinal model for the growth of the radiomic features is defined as: \[
y_{ij}^{(l)} = \mathbf{x}_i' \boldsymbol{\beta}^{(l)} + f^{(l)}(t_{ij}, \tau_i) + \mathbf{b}_i^{(l)} + \epsilon_{ij}^{(l)}
\]
for $l = 1, \dots, L$. Here, $\mathbf{x}_i$ represents the time-invariant demographic covariates, $\tau_i$ is the binary variable representing the treatment arm. The random effects $b_i^{(l)}$ vary across patients and follows the structure:\[
\mathbf{b}_i = \{\mathbf{b}_i^{(l)} : l = 1, \dots, L\}, \quad \text{where} \quad \mathbf{b}_i \sim \mathcal{N}_L(\mathbf{0}, \boldsymbol{\Sigma})
\]
The error terms $\epsilon_{ij}^{l}$ are uncorrelated Gaussian noise with variance $\sigma^2$. The function $f^l (t_{ij},\tau_i)$ captures the growth pattern of the longitudinal feature $l$, modeled as a natural spline function with time $t_{ij}$, affected by treatment $\tau_i$.

The survival component is based on the Cox proportional hazards model, where only the time-invariant demographic features are included: \[
h_i (t_{ij}) = h_0 (t_{ij}) \exp (\mathbf{x}_i' \boldsymbol{\beta}^{(h)})
\]
where $h_0 (t)$ is the baseline hazard, $h_i (t)$ is the hazard function for patient $i$ at time $t$, and $\boldsymbol{\beta}^{(h)}$ are the regression coefficients for the time-invariant covariates.

Combining both the longitudinal and survival components results in the joint component, where the hazard rate $\lambda_{ij}$, defined as the probability that the $i^{\text{th}}$ patient has an event at the $j^{\text{th}}$ observation time and not before, is given by:
\[
\text{logit} (\lambda_{ij}) = \mathbf{x}_i \boldsymbol{\beta}^{(h)} + f^h (t_{ij},\tau_i) + \sum_{l=1}^{L} \alpha_l^{(h)} \hat{y}_{ij}^{l}
\]
where $\hat{y}_{ij}^{l}$ are the predictions from the longitudinal components, and $\alpha_l^{(h)}$ are the coefficients (free association parameters) linking the longitudinal processes to the survival outcome. 

\subsection {Variable Selection for Joint Modeling Analysis}\label{variablejm}
To handle the high-dimensional radiomic feature space, we employed Best Subset Selection (BESS) to identify the most significant predictors at multiple time points (Weeks 24, 16, 8, and baseline). BESS was applied sequentially, selecting between 2 and 20 features at each timepoint to evaluate model performance with varying subset sizes. For each timepoint, the BESS algorithm was used in conjunction with a Cox proportional hazards model, and features selected in the best-performing models were recorded.

\begin{figure}[!htbp]
    \centering
    \includegraphics[width=1\textwidth]{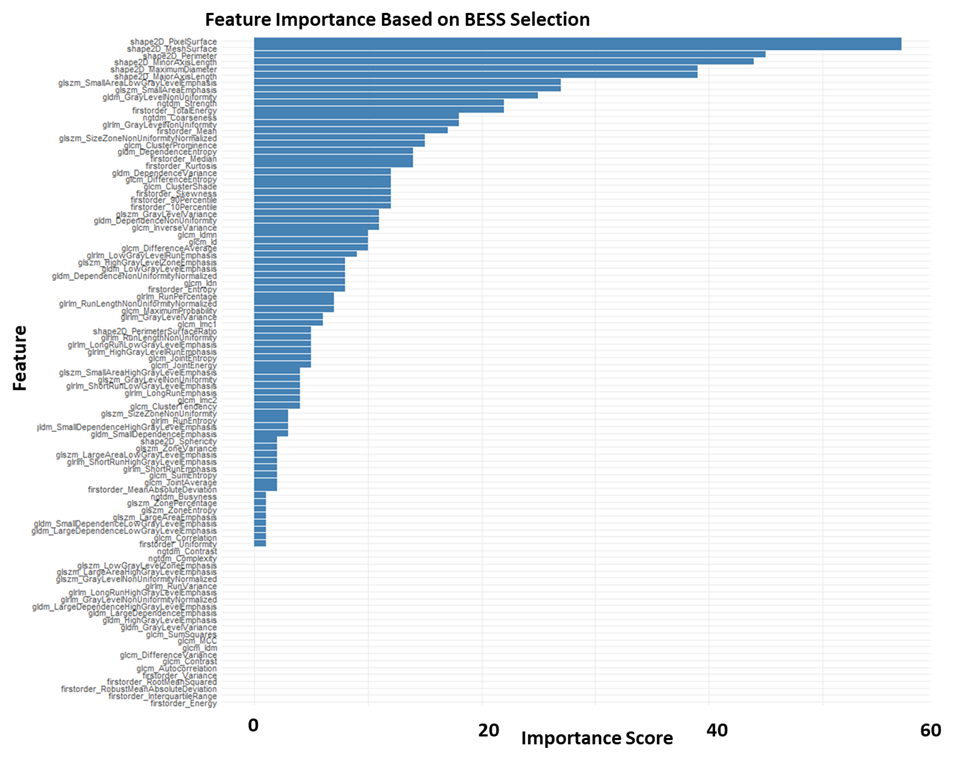} 
    \caption{Feature importance based upon best subset selection and voting procedure}
    \label{fig7}
\end{figure}

A scoring system was implemented, where each feature was assigned a score based on how frequently it appeared in the selected models across different subset sizes. The process was repeated for each timepoint, and the scores were accumulated, reflecting the overall significance of each feature throughout the longitudinal analysis. The features were subsequently ranked based on their total score, with higher-ranked features representing those consistently selected across time points, indicating their importance in predicting patient outcomes. Figure~\ref{fig7} presents the feature importance based on BESS selection, with the shape 2D radiomic features exhibiting the highest importance scores.

\subsection{Joint Modeling Results}\label{jmresults}

In Model 1, a joint model is established, combining both longitudinal radiomic data and survival data to explore the relationship between the top 3 Shape 2D-based radiomic features and progression-free survival (PFS) in metastatic breast cancer patients. The survival component uses a Cox proportional hazards model, while the longitudinal component consists of three linear mixed-effects models. The longitudinal models track changes in radiomic features (\textit{shape2D\_Elongation}, $\log$(\textit{shape2D\_MeshSurface}), and $\log$(\textit{shape2D\_MinorAxisLength})) over multiple time points (Baseline, Weeks 8, 16, and 24), with the treatment arm and baseline time-invariant demographic covariates (ALP, HGB, AST, AGE, LIVERBL) as fixed effects.

The joint modeling approach utilized a Markov Chain Monte Carlo (MCMC) algorithm \cite{rizopoulos2012joint} to estimate the posterior distributions of the model parameters. MCMC sampling is carried out with 20,000 iterations, with a burn-in period of 1,000 iterations and a thinning interval of 10. 

\begin{figure}[h]
    \centering
    \includegraphics[width=1\textwidth]{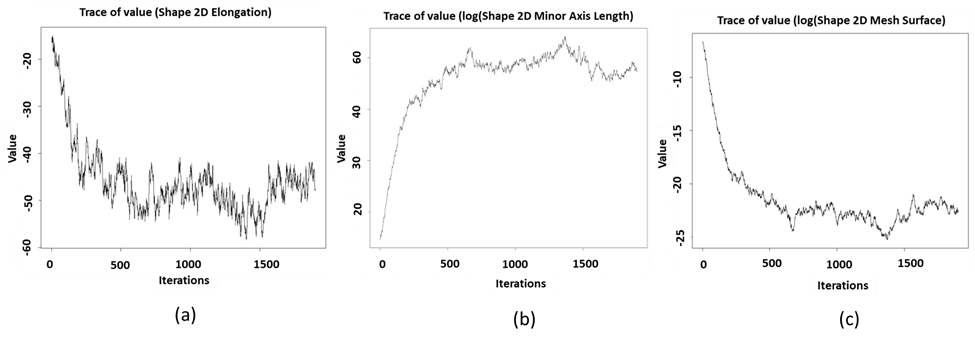} % Replace with actual filename
    \caption{Trace plots for (a) Shape 2D Elongation, (b) log(shape 2D Minor axis length) and (c) log (shape 2D mesh surface) shows to converge with iterations}
    \label{fig8}
\end{figure}

Trace plots in Figure~\ref{fig8} show that initially, the parameter values exhibit fluctuations, reflecting the adjustment phase of the sampling process (burn-in period). After this phase, the trace plots start to stabilize, with the parameter estimates fluctuating around a consistent mean, indicating that the model has converged after the specified number of iterations.

The summary of the joint model in table~\ref{tab1}, integrating longitudinal and survival outcomes, demonstrates that ribociclib treatment is significantly associated with a reduced hazard of progression ($\text{HR} = \exp(-2.5455)$, $p = 0.0053$). Patients with baseline liver metastasis exhibit a higher conditional risk of disease progression ($\text{HR} = \exp(3.6387)$, $p < 0.0001$). Longitudinal radiomic features, including \textit{shape2D Elongation}, $\log(\textit{Shape2D Mesh Surface})$, and $\log(\textit{Shape2D Minor Axis Length})$, are significantly associated with survival outcomes ($p < 0.0001$), indicating these features' predictive relevance over time. Additionally, a downward trend in these radiomic features was observed across time points, suggesting their dynamic role in disease progression. 

Model performance metrics, including \textit{Deviance Information Criterion (DIC)}, \textit{Watanabe-Akaike Information Criterion (WAIC)}, and \textit{Log Pseudo Marginal Likelihood (LPML)}, suggest a robust fit (see Table~\ref{tab2}).

\begin{table}[h!]
\centering
\caption{Summary table for longitudinal and survival outcomes for Model 1 based on training data}
\begin{tabular}{@{}lrrrrr@{}}
\toprule
\textbf{Parameter} & \textbf{Mean} & \textbf{Std. Dev.} & \textbf{2.5\% CI} & \textbf{97.5\% CI} & \textbf{p-value} \\ 
\midrule
\multicolumn{6}{l}{\textbf{Survival Outcome}} \\
ARM (Ribociclib)           & -2.55 & 1.23 & -4.95 & -0.40 & 0.0053 \\
LIVERBL Yes               & 3.64  & 0.92 & 1.56  & 5.12  & $<0.0001$ \\
Shape 2D Elongation       & -45.96 & 7.30 & -55.38 & -21.34 & $<0.0001$ \\
Shape 2D Mesh Surface     & -21.41 & 3.20 & -24.70 & -10.45 & $<0.0001$ \\
Shape 2D Minor Axis Length & 45.90 & 6.78 & 22.24 & 52.31 & $<0.0001$ \\
ALP                       & -0.01 & 0.01 & -0.03 & 0.01 & 0.0114 \\
HGB                       & -0.01 & 0.03 & -0.05 & 0.03 & 0.7421 \\
AST                       & 0.02  & 0.02 & -0.02 & 0.08 & 0.2937 \\
AGE                       & 0.05  & 0.03 & -0.01 & 0.10 & 0.1011 \\
\midrule
\multicolumn{6}{l}{\textbf{Longitudinal Outcome: Shape 2D Elongation}} \\
Intercept                 & 0.69 & 0.08 & 0.52 & 0.85 & $<0.0001$ \\
ns(Timepoint,2)[1]        & -0.09 & 0.02 & -0.12 & -0.05 & $<0.0001$ \\
ns(Timepoint,2)[2]        & -0.03 & 0.01 & -0.04 & -0.01 & $<0.0001$ \\
\midrule
\multicolumn{6}{l}{\textbf{Longitudinal Outcome: log(Shape 2D Mesh Surface)}} \\
Intercept                 & 5.04 & 0.52 & 4.01 & 6.03 & $<0.0001$ \\
ns(Timepoint,2)[1]        & -0.69 & 0.05 & -0.80 & -0.58 & $<0.0001$ \\
ns(Timepoint,2)[2]        & -0.36 & 0.04 & -0.43 & -0.30 & $<0.0001$ \\
\midrule
\multicolumn{6}{l}{\textbf{Longitudinal Outcome: log(Shape 2D Minor Axis Length)}} \\
Intercept                 & 2.36 & 0.27 & 1.84 & 2.88 & $<0.0001$ \\
ns(Timepoint,2)[1]        & -0.27 & 0.03 & -0.33 & -0.19 & $<0.0001$ \\
ns(Timepoint,2)[2]        & -0.17 & 0.01 & -0.21 & -0.15 & $<0.0001$ \\
\bottomrule
\end{tabular}
\label{tab1}
\end{table}

We developed two additional joint models (Model 2 and Model 3) to compare against Model 1 and evaluate the performance based on key statistical metrics such as \textit{DIC (Deviance Information Criterion)}, \textit{WAIC (Watanabe-Akaike Information Criterion)}, and \textit{LPML (Log Pseudo Marginal Likelihood)}. In Model 2, three radiomic features based on first-order statistics (\textit{First Order Total Energy}, \textit{First Order Mean}, \textit{First Order Kurtosis}) are used, which also demonstrated similar trends in terms of predictive utility. In Model 3, three radiomic features based on shape (\(\log(\textit{Shape 2D Maximum Diameter})\), \(\log(\textit{Shape 2D Major Axis Length})\), \(\log(\textit{Shape 2D Minor Axis Length})\)) are integrated, which exhibit high correlation with each other.
Table~\ref{tab2} shows that Model 1 demonstrates the best performance among the three models, with lowest DIC, lowest WAIC and highest LPML values. Model 3 has the poorest performance as it suffers from multicollinearity among the radiomic features. 

\begin{table}[h!]
\centering
\caption{Model performance based on DIC, WAIC and LPML criteria}
\begin{tabular}{@{}lrrr@{}}
\toprule
\textbf{Model} & \textbf{DIC} & \textbf{WAIC} & \textbf{LPML} \\ 
\midrule
Model 1 & 2031.46  & 6.1088e+03 & -5036.411 \\
Model 2 & 30149.21 & 3.3965e+04 & -17011.90 \\
Model 3 & 8065.19  & 8.5218e+08 & -755188.14 \\
\bottomrule
\end{tabular}
\label{tab2}
\end{table}

\section{Conclusion}\label{conclusion}

In this study we developed and applied comprehensive methodologies to analyze longitudinal radiomic data from mBC patients enrolled in two large-scale clinical trials, MONALEESA-3 and MONALEESA-7. The RAMAC algorithm allowed systematic and consistent tracking of target lesions across multiple CT time points (4 time points were used in the analysis), ensuring consistent lesion matching and accurate feature extraction. The integration of baseline radiomic features and delta radiomic across different post-treatment time points enabled us to develop predictive models for PFS that exhibited stable performance across longitudinal imaging time points.

We addressed the challenge of high-dimensional radiomic data by employing feature reduction techniques, including regularized additive Cox proportional hazards models and BESS, to mitigate the risk of overfitting and improve model interpretability. The results demonstrated that incorporating delta radiomic features (where the ratio of features at each timepoint are considered with respect to screening) across multiple time points, significantly improved the predictive power of the models, as evidenced by an increase in the C-index. With the increase in the information content compared to baseline the mean C-index increased from 0.58 using only the baseline radiomic features to 0.64 when delta radiomic features were added at later timepoints in the additive framework.

Furthermore, we utilized a JM approach, which allowed for the effective assessment of dynamic associations between longitudinal radiomic features and survival outcome. The JM model revealed significant temporal changes in radiomic marker. This study underscores the utility of longitudinal radiomic data for survival prediction in mBC.

\section*{Acknowledgement(s)}

This work is supported by the FDA Office of Women’s Health. Subrata Mukherjee’s appointment was supported by the Research Participation Program at the U.S. Food and Drug Administration administered by the Oak Ridge Institute for Science and Education through an interagency agreement between the U.S. Department of Energy and the U.S. Food and Drug Administration.

\section*{Disclaimer}

This article reflects the views of the authors and does not represent the views or policy of the U.S. Food and Drug Administration, the Department of Health and Human Services, or the U.S. Government.  The mention of commercial products, their sources, or their use in connection with material reported herein is not to be construed as either an actual or implied endorsement of such products by the Department of Health and Human Services.

\bibliographystyle{plainnat}
\bibliography{article}

\end{document}